\documentclass[twocolumn,letterpaper,aps,prc,superscriptaddress,showpacs,floatfix,preprintnumbers,longbibliography]{revtex4-1}

\usepackage{graphicx}   
\usepackage{xspace}     
\usepackage{url}
\usepackage[mathlines]{lineno}
\usepackage{bm}
\usepackage{amsmath}
\usepackage[dvipdfm]{hyperref}
\usepackage{xcolor}
\hypersetup{
	colorlinks = true,
	citecolor= blue,
	linkbordercolor = cyan
}
\newcommand{\pt}{\mbox{$p_T$}\xspace}

\newcommand{\sqsn}{\mbox{$\sqrt{s_{_{NN}}}$}\xspace}

\newcommand{\lam}{\mbox{$\Lambda$}\xspace}
\newcommand{\alam}{\mbox{$\bar{\Lambda}$}\xspace}
\usepackage{color}
\definecolor{orange}{cmyk}{0.,0.353,1.,0.}    

\begin{document}
%

\title{Global polarization of $\Lambda$ hyperons in Au+Au collisions
  at \sqsn = 200 GeV}

\affiliation{AGH University of Science and Technology, FPACS, Cracow 30-059, Poland}
\affiliation{Argonne National Laboratory, Argonne, Illinois 60439}
\affiliation{Brookhaven National Laboratory, Upton, New York 11973}
\affiliation{University of California, Berkeley, California 94720}
\affiliation{University of California, Davis, California 95616}
\affiliation{University of California, Los Angeles, California 90095}
\affiliation{Central China Normal University, Wuhan, Hubei 430079}
\affiliation{University of Illinois at Chicago, Chicago, Illinois 60607}
\affiliation{Creighton University, Omaha, Nebraska 68178}
\affiliation{Czech Technical University in Prague, FNSPE, Prague, 115 19, Czech Republic}
\affiliation{Nuclear Physics Institute AS CR, Prague 250 68, Czech Republic}
\affiliation{Technische Universitat Darmstadt, Darmstadt 64289, Germany}
\affiliation{Frankfurt Institute for Advanced Studies FIAS, Frankfurt 60438, Germany}
\affiliation{Fudan University, Shanghai, 200433 China}
\affiliation{Institute of Physics, Bhubaneswar 751005, India}
\affiliation{Indiana University, Bloomington, Indiana 47408}
\affiliation{Alikhanov Institute for Theoretical and Experimental Physics, Moscow 117218, Russia}
\affiliation{University of Jammu, Jammu 180001, India}
\affiliation{Joint Institute for Nuclear Research, Dubna, 141 980, Russia}
\affiliation{Kent State University, Kent, Ohio 44242}
\affiliation{University of Kentucky, Lexington, Kentucky 40506-0055}
\affiliation{Lamar University, Physics Department, Beaumont, Texas 77710}
\affiliation{Institute of Modern Physics, Chinese Academy of Sciences, Lanzhou, Gansu 730000}
\affiliation{Lawrence Berkeley National Laboratory, Berkeley, California 94720}
\affiliation{Lehigh University, Bethlehem, Pennsylvania 18015}
\affiliation{Max-Planck-Institut fur Physik, Munich 80805, Germany}
\affiliation{Michigan State University, East Lansing, Michigan 48824}
\affiliation{National Research Nuclear University MEPhI, Moscow 115409, Russia}
\affiliation{National Institute of Science Education and Research, HBNI, Jatni 752050, India}
\affiliation{National Cheng Kung University, Tainan 70101 }
\affiliation{Ohio State University, Columbus, Ohio 43210}
\affiliation{Institute of Nuclear Physics PAN, Cracow 31-342, Poland}
\affiliation{Panjab University, Chandigarh 160014, India}
\affiliation{Pennsylvania State University, University Park, Pennsylvania 16802}
\affiliation{Institute of High Energy Physics, Protvino 142281, Russia}
\affiliation{Purdue University, West Lafayette, Indiana 47907}
\affiliation{Pusan National University, Pusan 46241, Korea}
\affiliation{Rice University, Houston, Texas 77251}
\affiliation{Rutgers University, Piscataway, New Jersey 08854}
\affiliation{Universidade de Sao Paulo, Sao Paulo, Brazil, 05314-970}
\affiliation{University of Science and Technology of China, Hefei, Anhui 230026}
\affiliation{Shandong University, Jinan, Shandong 250100}
\affiliation{Shanghai Institute of Applied Physics, Chinese Academy of Sciences, Shanghai 201800}
\affiliation{State University of New York, Stony Brook, New York 11794}
\affiliation{Temple University, Philadelphia, Pennsylvania 19122}
\affiliation{Texas A\&M University, College Station, Texas 77843}
\affiliation{University of Texas, Austin, Texas 78712}
\affiliation{University of Houston, Houston, Texas 77204}
\affiliation{Tsinghua University, Beijing 100084}
\affiliation{University of Tsukuba, Tsukuba, Ibaraki 305-8571, Japan}
\affiliation{Southern Connecticut State University, New Haven, Connecticut 06515}
\affiliation{University of California, Riverside, California 92521}
\affiliation{University of Heidelberg, Heidelberg, 69120, Germany }
\affiliation{United States Naval Academy, Annapolis, Maryland 21402}
\affiliation{Valparaiso University, Valparaiso, Indiana 46383}
\affiliation{Variable Energy Cyclotron Centre, Kolkata 700064, India}
\affiliation{Warsaw University of Technology, Warsaw 00-661, Poland}
\affiliation{Wayne State University, Detroit, Michigan 48201}
\affiliation{Yale University, New Haven, Connecticut 06520}

\author{J.~Adam}\affiliation{Creighton University, Omaha, Nebraska 68178}
\author{L.~Adamczyk}\affiliation{AGH University of Science and Technology, FPACS, Cracow 30-059, Poland}
\author{J.~R.~Adams}\affiliation{Ohio State University, Columbus, Ohio 43210}
\author{J.~K.~Adkins}\affiliation{University of Kentucky, Lexington, Kentucky 40506-0055}
\author{G.~Agakishiev}\affiliation{Joint Institute for Nuclear Research, Dubna, 141 980, Russia}
\author{M.~M.~Aggarwal}\affiliation{Panjab University, Chandigarh 160014, India}
\author{Z.~Ahammed}\affiliation{Variable Energy Cyclotron Centre, Kolkata 700064, India}
\author{N.~N.~Ajitanand}\affiliation{State University of New York, Stony Brook, New York 11794}
\author{I.~Alekseev}\affiliation{Alikhanov Institute for Theoretical and Experimental Physics, Moscow 117218, Russia}\affiliation{National Research Nuclear University MEPhI, Moscow 115409, Russia}
\author{D.~M.~Anderson}\affiliation{Texas A\&M University, College Station, Texas 77843}
% ...

\collaboration{STAR Collaboration}\noaffiliation

\date{\today}

\begin{abstract} 
Global polarization of \lam hyperons has been measured to be of the
order of a few tenths of a percent in Au+Au collisions at \sqsn = 200
GeV, with no significant difference between \lam and \alam.  These new
results reveal the collision energy dependence of the global polarization
together with the results previously observed at \sqsn = 7.7 -- 62.4
GeV and indicate noticeable vorticity of the medium created in
non-central heavy-ion collisions at the highest RHIC collision
energy. The signal is in rough quantitative agreement with the
theoretical predictions from a hydrodynamic model and from the AMPT (A
Multi-Phase Transport) model.  The polarization is larger in more
peripheral collisions, and depends weakly on the hyperon's transverse
momentum and pseudorapidity $\eta^H$ within $|\eta^H|<1$.  An
indication of the polarization dependence on the event-by-event charge
asymmetry is observed at the $2\sigma$ level, suggesting a possible
contribution to the polarization from the axial current induced by the
initial magnetic field.
\end{abstract}

\maketitle

\setlength\linenumbersep{0.10cm}
\section{Introduction\label{sec:intro}}
Nucleus-nucleus collisions at the Relativistic Heavy Ion Collider and
at the Large Hadron Collider produce a state of partonic matter, the
Quark-Gluon Plasma (QGP), that is expected to have existed in nature
right after the Big Bang~\cite{Yagi:2005yb}.  Various experimental observations
together with sophisticated theoretical calculations indicate that
the QGP behaves as a nearly perfect liquid, i.e. a fluid with the
lowest ratio of shear viscosity to entropy density 
($\eta/s$)~\cite{Heinz:2013th,Voloshin:2008dg,schenke}.

One of the most important observables in heavy-ion experiments is the
azimuthal anisotropic flow that is usually quantified by the Fourier
coefficients of the azimuthal distribution of the final-state
particles relative to the collision symmetry planes. The first-order
coefficient, called the directed flow, is argued to be sensitive to the
equation of state of the matter, and could serve as a possible
signature of the QGP phase
transition~\cite{v1_wiggle,v1_eos,BES_pv1}. The second-order 
coefficient, elliptic flow, offers strong evidence for
the fluid-like behavior of the created matter. Furthermore, the
higher-order coefficients are found to provide additional constraints on
$\eta/s$ and the initial conditions.  In spite of a successful
description of the flow observables for $n\geq 2$ by hydrodynamic
models, none of the theoretical models can describe quantitatively 
the directed flow. This indicates that the current models still lack
an important ingredient in the description of relativistic
heavy-ion collision dynamics. The initial condition in the longitudinal 
direction would play an important role for the directed flow and 
vorticity~\cite{Bozek:2010bi,Becattini:2015ska}.

Several theoretical models suggest that the large angular momentum
carried by two colliding nuclei~\cite{Liang_2005,Voloshin:2004ha,Becattini_2008} 
can be transferred to the created system. As a consequence, the spin of
particles composing the system might be globally polarized along the
direction of the system angular momentum, due to spin-orbit
coupling. Such a global polarization can be measured experimentally
with hyperons via parity-violating weak decays, in which the daughter
baryon is preferentially emitted in the direction of the hyperon spin.
If the parent hyperon is an antiparticle, the daughter baryon tends
to be emitted in the opposite direction to the parent spin.

The angular distribution of daughter baryons in the hyperon decays is
given by
\begin{eqnarray}
\frac{dN}{d\cos\theta^{\ast}} \propto 1+\alpha_H P_{H}\cos\theta^{\ast},
\end{eqnarray}
where $\alpha_H$ is the hyperon decay constant, $P_H$ is the hyperon
polarization, and $\theta^{\ast}$ is the angle between the momentum of
daughter baryon and the polarization vector in the hyperon rest frame.
Since the angular momentum of the system is perpendicular to
the reaction plane (a plane defined by the impact
parameter vector and the beam direction), the polarization of hyperons
can be measured via the azimuthal distribution of daughter baryons 
with respect to the reaction plane in the hyperon rest frame, similar to anisotropic flow
measurements~\cite{Voloshin:2008dg}.

The STAR Collaboration performed the first global polarization
measurements of \lam hyperons in Au+Au collisions at \sqsn = 62.4 and
200 GeV in 2007~\cite{pol2007}. These results were consistent with
zero within large statistical uncertainties. More recently, the STAR
Collaboration has reported a non-zero signal for the \lam global
polarization in Au+Au collisions at lower energies (\sqsn = 7.7-39
GeV)~\cite{polBES}, with a possible difference between \lam and \alam
polarizations that may indicate the effect of the spin alignment by
the initial magnetic field. These results can be qualitatively
described by hydrodynamic and transport
models~\cite{polHydro,polAMPT}. The global polarization seems to
decrease with increasing collision energy, and those models predict a
finite signal ($\sim$0.2\%) at the top RHIC energy, \sqsn = 200 GeV.
It is thus important to measure the global polarization signal at
\sqsn = 200 GeV with all available statistics, in order to enhance
understanding of the role of vorticity in heavy-ion collisions. It is
likely related to other observables such as directed flow, elliptic
flow, and the source tilt of the system measured via
femtoscopy~\cite{Becattini_2008,Becattini:2015ska,Lisa:2000xj}.
Ref.~\cite{Baznat:2017jfj} explains the observed global polarization
as a result of the axial charge separation due to the Chiral Vortical
Effect.   Similar to the
Chiral Magnetic Effect, which is the induction of an electric current
along the magnetic field in a medium with non-zero axial charge, an
axial current can be generated in the medium with non-zero baryon
chemical potential by the system vorticity via the Chiral Vortical
Effect (for a recent review of the chiral anomalous effects in heavy-ion collisions, 
see ~\cite{Kharzeev:2015znc}). 
Thus the global polarization measurements might provide important 
information on the chiral dynamics of the system. 
Furthermore, precise measurements of
the difference in the polarization between \lam and \alam provide
constraints on the magnitude and the lifetime of the magnetic field in
heavy-ion collisions~\cite{Becattini:2016gvu}.

In this paper, we present results of the global polarization of
\lam and \alam hyperons in Au+Au collisions at \sqsn = 200 GeV using
the data recorded by the STAR experiment in the years 2010, 2011, and
2014.  The total dataset is about 150 times
larger than the dataset analyzed in the previous search by STAR for
hyperon polarization in Au+Au collisions at \sqsn = 200
GeV~\cite{pol2007}.  We present the results as functions of the
collision centrality, the hyperon's transverse momentum, and
pseudorapidity. We also present comparisons with available theoretical
calculations. Furthermore, we present the dependence of the
polarization on the event-by-event charge asymmetry, to study a
possible relation between the polarization and axial current induced
by the initial magnetic field~\cite{Voloshin:2017kqp}.

\section{STAR Experiment\label{sec:exp}}

The STAR detector is composed of central barrel detectors used for
tracking and particle identification, and trigger detectors located in
the forward and backward directions~\cite{Ackermann:2002ad}.  
Charged tracks were measured using the time projection chamber (TPC)~\cite{tpc}, 
which covers the full azimuth and a pseudorapidity range of $-1<\eta<1$. 
Momenta of charged particles were determined via trajectories of
reconstructed tracks and a primary vertex was reconstructed by
extrapolating the tracks back to the origin.  The TPC also allows 
particle identification based on the ionization energy loss, $dE/dx$, in
the TPC gas (Ar~90\% + CH$_{4}$~10\%).

The time-of-flight detector (TOF)~\cite{tof} is installed outside
the TPC, covering the full azimuth and a pseudorapidity range of
$-0.9<\eta<0.9$. Multi-gap resistive plate chamber (MRPC) technology 
is employed for the STAR TOF detector.
The TOF system consists of 120 trays and each tray has 32 MRPCs.
The timing resolution of the TOF system with a start
time from the vertex position detectors (VPD)~\cite{vpd} is
$\sim$100~ps. The TOF extends the capability of particle
identification provided by the TPC up to $p_{T}=3$ GeV/$c$.

The Zero Degree Calorimeters (ZDC)~\cite{zdc} and the VPD were used to
determine a minimum-bias trigger.  The ZDCs are located at forward (west) 
and backward (east) angles, $|\eta|>6.3$. The ZDCs are Cherenkov-light 
sampling calorimeters and each ZDC is composed of three identical modules.
They measure the energy deposit of spectator neutrons. 
The VPD consists of two identical sets of
detectors located at forward and backward rapidities and surrounds
the beam pipe, covering a pseudorapidity range of $4.24<|\eta|<5.1$.
Each VPD consists of nineteen modules, which is composed of a plastic scintillator 
with a Pb converter. The VPD also provides the start time of collisions and 
the position of the collision vertex along the beam direction.

\section{Data Analysis\label{sec:ana}}

The analysis is based on the data for Au+Au collisions at \sqsn =
200~GeV taken in the years 2010, 2011, and 2014 with a minimum-bias
trigger selected by a coincidence signal between the east and west VPDs. 
The collision vertex along the beam direction was required to
be within 30~cm of the center of the TPC for 2010 and 2011 data and to
be within 6~cm for 2014 data. In the 2014 data the narrower vertex
selection was required to ensure a good acceptance 
for the Heavy Flavor Tracker (HFT) installed prior to 2014
run~\cite{Schambach:2014uaa,Contin:2016hpt} (Note that the HFT was not used in this analysis).  
Additionally, the difference between the vertex positions along the beam direction
determined by the TPC and the VPD was required to be less than 3~cm,
to reduce the beam-induced background.  The vertex position in the
transverse plane was limited to be within 2~cm from the beam
line. These selection criteria yielded two hundred million events
using the 2010 dataset, three hundred fifty million events using the
2011 dataset, and one billion events using the 2014 dataset.  The
collision centrality was determined based on the measured multiplicity
of charged tracks within $|\eta|<0.5$, and this was matched to a Monte
Carlo Glauber simulation in the same way as in previous
studies~\cite{BESv2}. The effect of the trigger efficiency was taken
into account in the analysis by weighting events especially in peripheral collisions 
when calculating final results, although the effect is very small.

\subsection{Event plane determination}\label{sec:EP}

\begin{figure}[hbt]
\begin{center}
\includegraphics[width=0.9\linewidth]{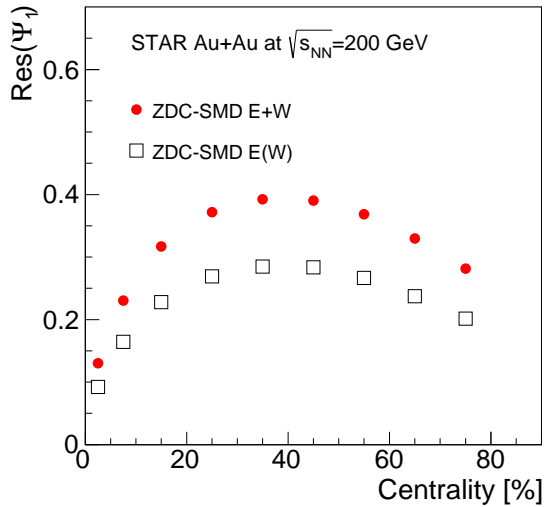}
\caption{Resolution of the first-order event plane determined by the
  ZDC-SMDs~\cite{zdc} in Au+Au collisions at \sqsn = 200 GeV; ZDC-SMD E+W
  denotes the combined plane of ZDC-SMDs in east and west sides and
  ZDC-SMD E(W) denotes one of the ZDC-SMDs.\label{fig:resZDC}}
\end{center}
\end{figure}

As an experimental estimate of the reaction plane, the first-order
event plane $\Psi_1$ was determined by the ZDCs that are equipped with Shower
Maximum Detectors (SMD)~\cite{zdc,v1smd,v1smd_star}. 
The ZDCs measure the energy deposited by spectator neutrons, and the
SMDs measure the centroid of the hadronic shower caused by the
interaction between spectator neutrons and the ZDC. Since the
spectator neutrons are deflected outward from the centerline of the
collisions~\cite{spflow}, we can determine the direction of the
angular momentum of the system (see Ref.~\cite{Adamczyk:2017ird} for
more details). The event plane resolution, 
Res($\Psi_1$) = $\langle\cos(\Psi_1-\Psi_1^{\rm obs})\rangle$,
was estimated by the two-subevent method~\cite{TwoSub}, 
where $\Psi_1^{\rm obs}$ denotes a measured event plane. 
Figure~\ref{fig:resZDC} shows the
event plane resolution for the year 2011 data as an example.  The
resolution reaches a maximum of $\sim$0.39 around 30\%-40\% centrality
for the combined plane of ZDC-SMD east and west.  The resolution is
consistent between 2010 and 2011 data and is better by $\sim$5\% for
2014 data compared to that for 2011.

\subsection{Track selection}

Charged tracks reconstructed from the TPC hit information were
selected with the following requirements to assure good quality. The
number of hit points used in the reconstruction was required to be
greater than 14. The ratio of the number of hit points used to the
maximum possible number of hit points (45 for a track traversing the
entire TPC, but the maximum number can be smaller when track trajectory 
reaches an endcap of the TPC) was required to be larger than~0.52. 
Tracks corresponding to $0.15<\pt<10$~GeV/$c$ and $|\eta|<1$ were used 
in this study.

\subsection{\lam reconstruction\label{sec:lamreco}}

\lam hyperons were identified via decay channels
$\lam\rightarrow p+\pi^{-}$ and
$\alam\rightarrow \bar{p}+\pi^{+}$. These decay modes account for
(63.9$\pm$0.5)\% of all decays~\cite{PDG}. The daughter particles of
\lam and \alam, i.e. charged pions and protons, were identified by
using $dE/dx$ information from the TPC and time-of-flight information
from the TOF detector, like in our previous
publication~\cite{Adamczyk:2017ird}. Charged pions and protons were
selected by requiring the track to be within three standard deviations
(3$\sigma$) from their peaks in the normalized $dE/dx$
distribution. If the track had TOF hit information, a constraint based
on the square of the measured mass was required.  If the TOF information was
not available, an additional cut based on $dE/dx$ was applied, 
requiring pions (protons) to be 3$\sigma$ away from the proton (pion)
peak in the normalized $dE/dx$ distribution.

The invariant mass, $M_{\rm inv}$, was calculated using candidates for the daughter
tracks. To reduce the combinatorial background, selection criteria
based on the following decay topology parameters were used:
\begin{itemize}
\item Distance of the closest approach (DCA) between daughter tracks
  and the primary vertex,
\item DCA between reconstructed trajectories of \lam (\alam)
  candidates and the primary vertex,
\item DCA between two daughter tracks,
\item Decay length of \lam (\alam) candidates.
\end{itemize}
Furthermore \lam (\alam) candidates were required to point away from the primary vertex.
Cuts on the decay topology were adjusted, depending on the
collision centrality, to account for the variation of the combinatorial
background with centrality. The background level relative to the \lam
(\alam) signal in the $\Lambda$ mass region falls below 30\% at maximum in this analysis.  
Finally, \lam and \alam with $0.5<\pt<6$~GeV/$c$ and $|\eta|<1$ were
analyzed in this study.

\begin{figure}[hbt]
\begin{center}
\includegraphics[width=\linewidth]{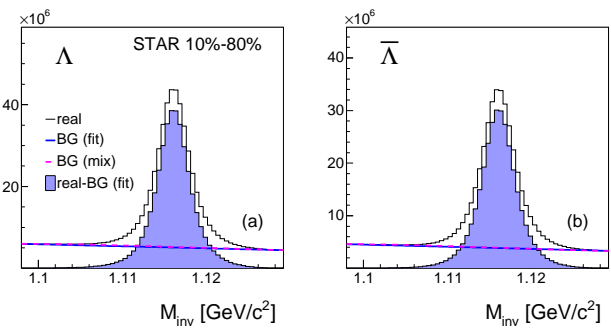}
\caption{Invariant mass distributions of the ($p$,\,$\pi^{-}$) system for \lam (a)
  and of the ($\bar{p}$,\,$\pi^{+}$) system for \alam (b) in the 30-40\% centrality
  bin for 2014 data. Bold solid lines show the background distribution obtained by a
  linear fitting function, and dashed lines show the background from mixed events. 
  Shaded areas show the extracted signal after the background subtraction
  using the fitting function.\label{fig:IM}}
\end{center}
\end{figure}

Figure~\ref{fig:IM} shows the invariant mass distributions for \lam
and \alam in the 10\%-80\% centrality bin for 2014 data as an example. The combinatorial
background under the \lam~peak was estimated by fitting the off-peak region
with a linear function, and by the event mixing technique~\cite{Adamczyk:2013gw}, 
shown in Fig.~\ref{fig:IM} as solid and dashed lines, respectively.

\subsection{Polarization measurement}\label{sec:pol}

As mentioned in Sec.~\ref{sec:intro}, the global polarization can be
measured via analysis of the azimuthal distribution of daughter
protons in the \lam rest frame relative to the reaction plane. As
mentioned in Sec.~\ref{sec:EP}, the first-order event plane $\Psi_1$
determined by the spectator fragments was used in this analysis as an
estimator of the reaction plane.  The sideward deflection of the
spectators allows us to know the direction of the initial angular
momentum.  Taking into account the experimental resolution of the
event plane, the polarization projected onto the direction of the
system global angular momentum can be obtained by~\cite{pol2007}:
\begin{eqnarray}
P_{H} = \frac{8}{\pi\alpha_{H}} \frac{\langle\sin(\Psi_{1}^{\rm obs}-\phi_{p}^{\ast})\rangle}{\rm Res(\Psi_1)},
\label{eq:PHep}
\end{eqnarray}
where $\alpha_{H}$ are the decay parameters of \lam ($\alpha_{\Lambda}$)
and \alam ($\alpha_{\bar{\Lambda}}$), $\alpha_{\Lambda}=-\alpha_{\bar{\Lambda}}=0.642\pm0.013$~\cite{PDG}. 
The angle $\phi_{p}^{\ast}$ denotes the azimuthal angle of the daughter
proton in the \lam rest frame.  The Res($\Psi_1$) is the resolution of
the first-order event plane.  Two different techniques were used to
extract the polarization signal
$\langle\sin(\Psi_{1}-\phi_{p}^{\ast})\rangle$: the invariant mass
method and the event plane method, both of which are often used in 
flow analyses~\cite{Borghini:2004ra,Voloshin:2008dg}.

\begin{figure}[hbt]
\begin{center}
\includegraphics[width=\linewidth]{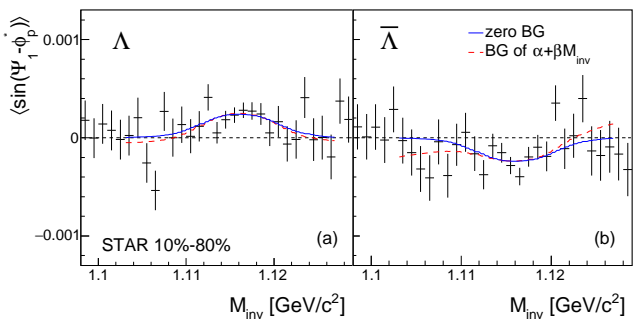}
\caption{\label{fig:SinIM}$\langle
  \sin(\Psi_1-\phi_{p}^{\ast})\rangle$ as a function of the invariant
  mass for \lam (a) and \alam (b) in the 10\%-80\% centrality bin for 2014 data. Solid and
  dashed lines show the fitting function for actual fit range, Eq.~\ref{eq:SinIM}, with
  two different background assumptions.}
\end{center}
\end{figure}

In the invariant mass method~\cite{Borghini:2004ra,Adamczyk:2013gw}, 
the mean value of the sine term in
Eq.~(\ref{eq:PHep}) was measured as a function of the invariant
mass. Since the \lam particles and background cannot be separated on
an event-by-event basis, the observed polarization signal is the sum
of the signal and background:
\begin{eqnarray}
\langle\sin(\Psi_{1}-\phi_{p}^{\ast})\rangle^{\rm obs} &= (1-f^{\rm Bg}(M_{\rm
  inv}))\langle\sin(\Psi_{1}-\phi_{p}^{\ast})\rangle^{\rm Sg} \nonumber
\\ &{~~}+ f^{\rm Bg}(M_{\rm
  inv})\langle\sin(\Psi_{1}-\phi_{p}^{\ast})\rangle^{\rm Bg}, \label{eq:SinIM}
\end{eqnarray}
where $f^{\rm Bg}(M_{\rm inv})$ is the background fraction at the invariant mass
$M_{\rm inv}$.  The term $\langle\sin(\Psi_{1}-\phi_{p}^{\ast})\rangle^{\rm Sg}$ is
the polarization signal for \lam (\alam), where the term
$\langle\sin(\Psi_{1}-\phi_{p}^{\ast})\rangle^{\rm Bg}$ is the background
contribution, which is in general expected to be zero, but could be
non-zero, for example, due to misidentification of particles or errors
in track reconstruction.  The data were fitted with
Eq.~(\ref{eq:SinIM}) to extract the polarization signal.  Since the
shape of the background as a function of invariant mass is unknown,
two assumptions concerning the background contribution were tested: a
linear function over $M_{\rm inv}$
($\langle\sin(\Psi_{1}-\phi_{p}^{\ast})\rangle^{\rm Bg}=\alpha+\beta M_{\rm inv}$)
and zero background contribution ($\alpha=0$, $\beta=0$).
Figure~\ref{fig:SinIM} shows the observed $\langle
\sin(\Psi_{1}-\phi_{p}^{\ast})\rangle$ as a function of the invariant mass $M_{\rm
  inv}$.  Since the daughter proton tends to be emitted in the
direction of the parent hyperon spin, and in the opposite direction
for antiparticles, the $\langle \sin(\Psi_{1}-\phi_{p}^{\ast})\rangle$ for \alam
shows negative values around its mass region as shown in
Fig.~\ref{fig:SinIM}(b), while it is positive for \lam as in
Fig.~\ref{fig:SinIM}(a).  We found that results from these two
background assumptions give consistent results within uncertainties,
and the difference was incorporated in the systematic uncertainty as
described in the following section.

Although the invariant mass method was used as the default method in this analysis,  
the event plane method was also tested as a systematic check. 
In the event plane method, the same procedure as used in flow
analyses was utilized~\cite{Voloshin:2008dg}. First, the number of \lam and \alam was counted
in each bin of the hyperon emission azimuthal angle relative to the
event plane after the background subtraction, as demonstrated in
Fig.~\ref{fig:IM}.  Then the yield of \lam and \alam as a function of
$\Psi_{1}-\phi^{\ast}$ was fitted with a sine function to obtain the
mean sine $\langle\sin(\Psi_{1}-\phi_{p}^{\ast})\rangle^{\rm Sg}$.  The difference in 
results from the invariant mass and event plane methods  
is included in the systematic uncertainty.

\subsection{Effect of feed-down}
A sizable number of \lam and \alam produced in the collisions are
secondary particles -- products of heavier particle decays, such as
$\Sigma^{\ast}\rightarrow\lam+\pi$, $\Sigma^{0}\rightarrow\lam+\gamma$,
and $\Xi\rightarrow\lam+\pi$. The parent particles are also polarized.
The polarization is transferred from the parent particle to the 
daughter \lam. The contribution of such feed-down to the measured
polarization was studied in Refs. \cite{Becattini:2016gvu,polHydro,polAMPT}
and was found to dilute the polarization of the primary {\lam} by
15\%--20\%. Note that this estimate is model-dependent. In addition,
this effect might be smaller in our analysis due to reduction of
secondary particles by cuts on the decay topology of \lam. 
Below, the results are compared to models which do and do not 
take into account the feed-down effect.

\subsection{Systematic uncertainties}
The systematic uncertainties were estimated by varying topological
cuts, and comparing the results obtained with different methods for
the signal extraction and for the event plane determination.  Below we
describe each systematic source and provide typical values.

We applied ten different topological cuts and used the standard
deviation from the default cut set results as the symmetric systematic
uncertainty. The effect from the variation of the topological cuts was
found to be $<$3\%.

As described in the previous section, two different techniques were
used to extract the polarization signal. We used the result obtained
with the invariant mass method as default results, and the difference
in the results from the event plane method was included in the
systematic uncertainty. The difference in polarization based on different 
methods was found to be $\sim$21\%.

The first-order event plane determined by both ZDC-SMDs in the east
and west sides was used in this analysis.  For a cross check, the
event plane determined by each ZDC-SMD on its own was also used in the
analysis, although the poorer event-plane resolution resulted in
larger statistical uncertainties. The difference between the results
was included in the systematic uncertainty ($\sim$22\%).

According to Ref.~\cite{PDG}, the decay parameter for $\lam\rightarrow
p+\pi^{-}$, $\alpha_{\Lambda}$, is 0.642~$\pm$~0.013, while
$\alpha_{\bar{\Lambda}}=-0.71\pm 0.08$ for
\alam$\rightarrow\bar{p}+\pi^{+}$, based on world-average data.  If CP
is conserved, $\alpha_{\Lambda}=-\alpha_{\bar{\Lambda}}$.  In this
study, we use $\alpha_{\Lambda}=-\alpha_{\bar{\Lambda}}=0.642\pm0.013$
and the uncertainty in $\alpha_{H}$ was incorporated into the
systematic uncertainty ($\sim$2\%).  Also, the difference from the
case using $\alpha_{\bar{\Lambda}}=-0.71$, which we found to be
$\sim$9.6\%, was included in the systematic uncertainty for \alam.

As mentioned in Sec.\ref{sec:lamreco}, the combinatorial background in
the invariant mass distributions for \lam and \alam was estimated by a
linear function fit and by the event mixing technique as shown in
Fig.~\ref{fig:IM}. The difference between the results obtained with the 
two approaches was included in the systematic uncertainty ($<$1\%).

In the invariant mass method, the background contribution in the
off-peak region of \lam (\alam) mass distribution is unknown, but is
supposed to be zero as mentioned in Sec.~\ref{sec:pol}.  We confirmed
that the background signal was consistent with zero when increasing
the background by applying looser topological cuts.  Therefore, the
results from the zero-background assumption for the fitting function
were used as the final results, and the difference from the non-zero
background assumption was included in the systematic uncertainty
($\sim$13\%).

Final systematic uncertainties were calculated by taking the square
root of the quadratic sum of the difference between the default condition
and each systematic source.  We further examined whether or not there
is a possible experimental bias in our results. The data for Au+Au
collisions in the years 2010 and 2011 were taken with two different
polarities of the magnetic field. In order to check the effect of the
magnetic field configuration, we divided the data into two groups
according to the magnetic field polarity, and confirmed that there was
no significant difference between the two groups. Those two groups
also correspond to different times of data-taking. Despite changes in
the trigger conditions, which had the effect of further improving
data-taking during runs, and the associated change in the detector
conditions, no significant difference in the polarization results was
observed.

We also calculated the cumulant terms in a similar way as described  
in Ref.~\cite{Abelev:2008ag,Borghini:2002vp} 
and subtracted them from the observed signal to check for a possible 
detector effect due to non-uniformity in acceptance and a residual 
detector effect coming from the event plane calibration:
\begin{eqnarray}
\langle\langle\sin(\Psi_1-\phi_p^{\ast}) \rangle\rangle 
&-\langle\langle\sin\Psi_1\rangle\rangle
  \langle\langle\cos\phi_p^{\ast}\rangle\rangle \nonumber\\
&+\langle\langle\cos\Psi_1\rangle\rangle
  \langle\langle\sin\phi_p^{\ast}\rangle\rangle \label{eq:cumulant},
\end{eqnarray}
where the double angle brackets indicate an average over particles first, 
and then an average over events. It was found that the correction
terms are negligible and there was no significant difference in the
results beyond the current uncertainty due to the
correction. Therefore we did not apply this correction to the final
results.

The effect of the tracking efficiency was studied using a Geant simulation~\cite{Abelev:2008ag} 
and found to be negligible.  Also, the acceptance correction proposed in our previous
analysis~\cite{pol2007} was applied. The measured polarization can be
written as:
\begin{eqnarray}
\frac{8}{\pi\alpha_{H}} \langle\sin(\Psi_{\rm RP}-\phi_{p}^{\ast})\rangle =
A_0(p_T^H,\eta^H)P_H(p_T^H,\eta^H),
\end{eqnarray}
where $A_0$ is an acceptance correction factor defined as 
\begin{eqnarray}
A_0(p_T^H,\eta^H) &=&
\frac{4}{\pi}\langle\sin\theta_{p}^{\ast}\rangle \label{eq:A0}.
\end{eqnarray}
The correction factor $A_0$ was estimated using the experimental data.

The analysis was performed separately for each data set taken in
different years. As mentioned in Sec.~\ref{sec:EP}, the event plane resolution 
slightly differs in each year due to different detector conditions. Also, for 2014 data 
the tracking efficiency became worse at low \pt because of the HFT. We confirmed 
that this additional inefficiency does not affect our final results. Since the results from the years 2010, 2011,
and 2014 were consistent within their uncertainties, we
combined all results for the measured $P_{\rm H}$ to improve the statistical significance.


\section{Results\label{sec:result}}
%
\begin{figure}[htb]
\begin{center}
\includegraphics[width=\linewidth]{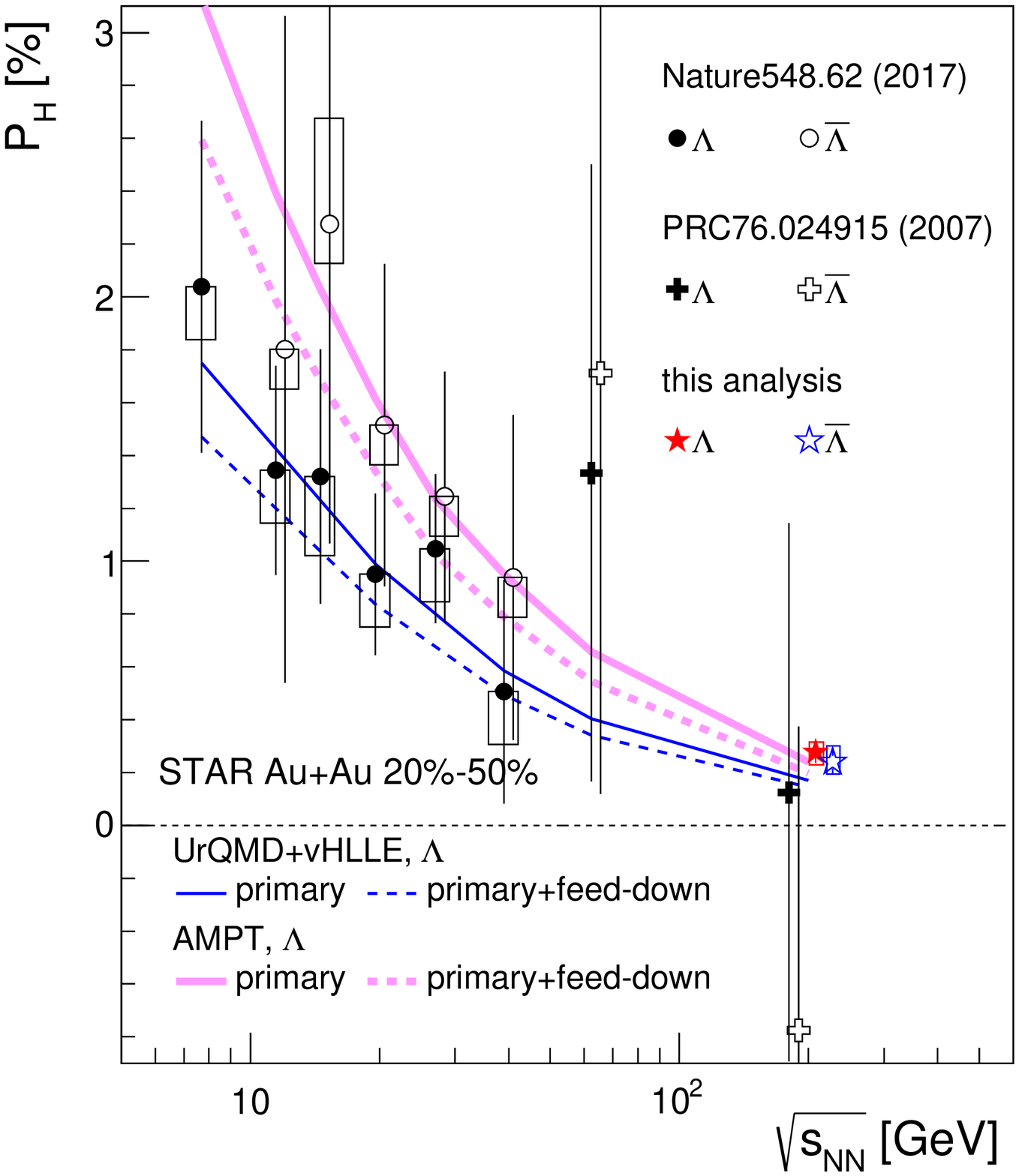}
\caption{\label{fig:PH_vsBES} Global polarization of \lam\ and \alam\ as
  a function of the collision energy \sqsn\ for 20-50\% centrality
  Au+Au collisions. Thin lines show calculations from
  a 3+1D cascade + viscous hydrodynamic model (UrQMD+vHLLE)~\cite{polHydro} 
  and bold lines show the AMPT model calculations~\cite{polAMPT}. In the case of each model,
  primary \lam\ with and without the feed-down effect are indicated by dashed and solid lines, respectively. 
  Open boxes and vertical lines
  show systematic and statistical uncertainties, respectively. Note
  that the data points at 200 GeV and for $\bar{\Lambda}$\ are slightly horizontally
  shifted for visibility.}
\end{center}
\end{figure}

Figure~\ref{fig:PH_vsBES} presents the global polarization of \lam and
\alam as a function of the collision energy for the 20--50\%
centrality bin in Au+Au collisions. The results from this analysis are
shown together with the results from lower collision energies \sqsn =
7.7--62.4 GeV~\cite{polBES}.  The 2007 result for \sqsn = 200 GeV~\cite{pol2007} has
a large uncertainty and is consistent with zero. Our new results for
\sqsn = 200 GeV with significantly improved statistical precision
reveal non-zero values of the polarization signal,
0.277~$\pm$~0.040~(stat)~$\pm$~$^{0.039}_{0.049}$~(sys) [\%] and
0.240~$\pm$~0.045~(stat)~$\pm$~$^{0.061}_{0.045}$~(sys) [\%] for \lam
and \alam respectively, and are found to follow the overall trend of
the collision energy dependence.  While the energy dependence of the
global polarization was not obvious from the lower energy results,
together with the new 200 GeV results, the polarization is found to
decrease at higher collision energy.  Calculations for primary \lam
and all \lam taking into account the effect of feed-down from a 3+1D
viscous hydrodynamic model vHLLE with the UrQMD initial
state~\cite{polHydro} are shown for comparison. The model calculations
agree with the data over a wide range of collision energies, including
\sqsn = 200 GeV within the current accuracy of our experimental
measurements. Calculations from a Multi-Phase Transport (AMPT) model
predict slightly higher polarization than the hydrodynamic model, but
are also in good agreement with the data within uncertainties. Neither
of the models accounts for the effect of the magnetic field or
predicts significant difference in \lam and \alam polarization due to
any other effect, e.g., non-zero baryon chemical potential makes the
polarization of particles lower than that of antiparticles, but the
effect is expected to be small~\cite{Fang:2016vpj}.  Other theoretical
calculations~\cite{Xie:2017upb,Baznat:2017jfj} such as a chiral
kinetic approach with the quark coalescence model~\cite{Sun:2017xhx}
can also qualitatively reproduce the experimental data.  It should be
noted that most of the models calculate the spin polarization from the
local vorticity at the freeze-out hypersurface. However it is not
clear when and how the vorticity and polarization are coupled during
the system evolution and how much the hadronic rescattering at the
later stage affects the spin polarization.

\begin{figure}[htb]
\begin{center}
\includegraphics[width=\linewidth]{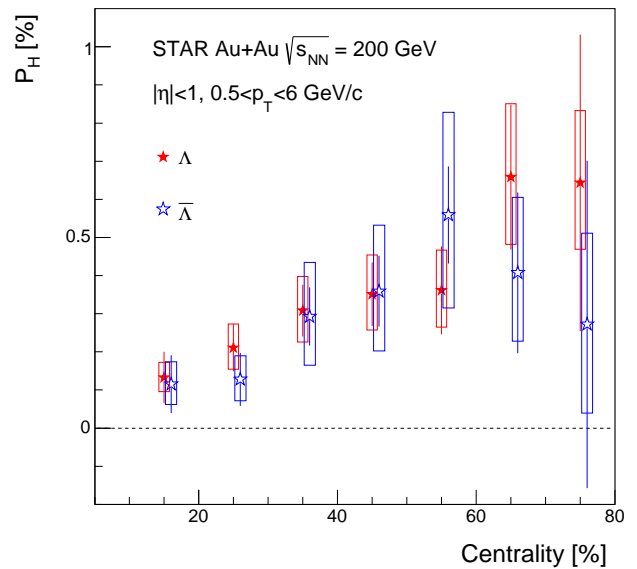}
\caption{\label{fig:PH_cent} \lam (\alam) polarization as a function
  of the collision centrality in Au+Au collisions at \sqsn = 200 GeV.
  Open boxes and vertical lines show systematic and statistical
  uncertainties.  The data points for $\bar{\Lambda}$ are slightly
  shifted for visibility.}
\end{center}
\end{figure}

We also performed differential measurements of the polarization, 
versus the collision centrality, the hyperon's transverse momentum, and
the hyperon's pseudorapidity. The vorticity of the system is expected to
be smaller in more central collisions because of smaller initial source
tilt~\cite{Bozek:2010bi,Adamczyk:2017ird}, and/or 
because the number of spectator nucleons becomes smaller. Therefore, 
the initial longitudinal flow velocity, which would be a source of the initial 
angular momentum of the system, becomes less dependent on the transverse
direction~\cite{Becattini_2008}.
Figure~\ref{fig:PH_cent} presents the centrality dependence of the
polarization.
The polarization of \lam and \alam is found to be larger in
more peripheral collisions, as expected from an increase in the thermal
vorticity~\cite{Jiang:2016woz}.  With the given large uncertainties, it is
not clear if the polarization saturates or even starts to drop off in the 
most peripheral collisions.

\begin{figure}[htb]
\begin{center}
\includegraphics[width=\linewidth]{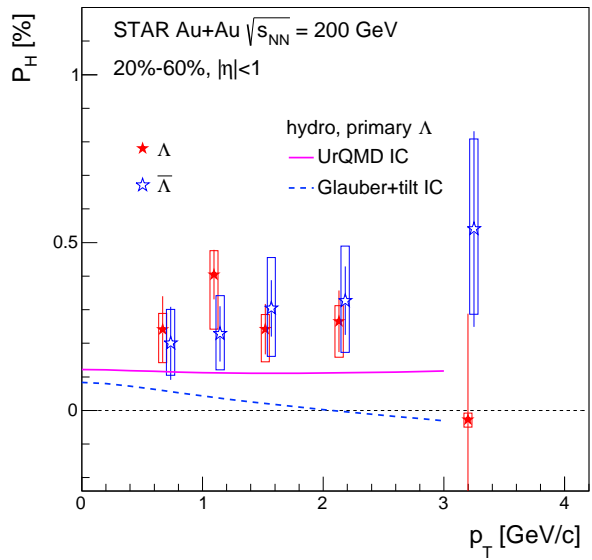}
\caption{\label{fig:PH_pt} Polarization of \lam and \alam as a
  function of \pt for the 20\%--60\% centrality bin in Au+Au collisions at
  \sqsn = 200 GeV.  Open boxes and vertical lines show systematic and statistical
  uncertainties, respectively. Hydrodynamic model calculations for \lam with
  two different initial conditions (IC) are compared.
  Note that the data points for $\bar{\Lambda}$ 
  are slightly shifted for visibility.}
\end{center}
\end{figure}

Figure~\ref{fig:PH_pt} shows the polarization as a function of \pt for
the 20\%--60\% centrality bin. The polarization dependence on \pt is weak
or absent, considering the large uncertainties, which is consistent with
the expectation that the polarization is generated by a rotation of
the system and therefore does not have a strong \pt dependence.  One
might expect a decrease of the polarization at lower \pt due to the 
smearing effect caused by scattering at the later stage of the
collisions, and/or a decrease of polarization at higher \pt because of a larger
contribution from jet fragmentation, but it is difficult to discuss
such effects given the current experimental uncertainties.  Calculations
for primary \lam from a hydrodynamic model with two different initial
conditions (ICs)~\cite{Becattini:2017gcx} are compared to the data.
The \pt dependence of the polarization slightly depends on the initial
conditions, i.e. Glauber IC with the initial tilt of the
source~\cite{Bozek:2010bi,Becattini:2015ska} and the initial state
from the UrQMD model~\cite{Karpenko:2015xea}.  The UrQMD IC includes
a pre-equilibrium phase which leads to the initial flow, but the Glauber
IC does not include it, and the initial energy density profile is
different between the two ICs, both of which would affect the initial
angular momentum.  The data are closer to the UrQMD IC, but on average
are slightly higher than the calculations.

\begin{figure}[htb]
\begin{center}
\includegraphics[width=\linewidth]{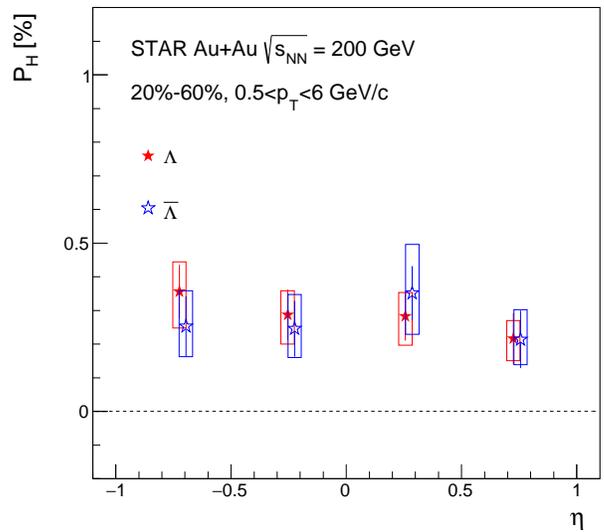}
\caption{\label{fig:PH_eta} Polarization of \lam and \alam as a
  function of $\eta$ for the 20\%--60\% centrality bin in Au+Au collisions at
 \sqsn = 200 GeV. Open boxes and vertical lines show systematic and statistical uncertainties.
  Note that the data points for $\bar{\Lambda}$ 
  are slightly shifted for visibility.}
\end{center}
\end{figure}

Figure~\ref{fig:PH_eta} presents the pseudorapidity dependence of the
polarization for \lam and \alam. It is consistent with being constant
within uncertainties. The vorticity is expected to decrease at large
rapidities, but might also have a local minimum at $\eta=0$ due to
complex shear flow
structure~\cite{polHydro,Jiang:2016woz,Deng:2016gyh}.  Due to baryon
transparency at higher collision energy and the event-by-event fluctuations
in the participant center-of-mass, such a dependence might be
difficult to observe within the acceptance of the STAR detector.

As mentioned in the introduction, the vorticity might be also related
to anomalous chiral effects~\cite{Kharzeev:2015znc}.  
In addition to the contribution from the Chiral Vortical Effect discussed 
in Ref.~\cite{Baznat:2017jfj}, the axial current ${\bf J}_{5}$ can be
generated in the medium with non-zero vector chemical potential
$\mu_{\rm v}$ by the magnetic field ${\bf B}$ (${\bf J}_{5}\propto e
\mu_{\rm v}{\bf B}$) via the Chiral Separation Effect~\cite{Soeren}.  Note that
${\bf J}_{5}$ points along the magnetic field in the case of $e
\mu_{\rm v}>0$ (where $e$ is the particle electric charge), but is
opposite for $e \mu_{\rm v}<0$.  Since the directions of the magnetic
field and the initial angular momentum of the system are parallel, an
additional contribution by ${\bf J}_5$ to the polarization might be
observed, i.e., for $e\mu_{\rm v}>0$ ($e\mu_{\rm v}<0$), the spins of
particles (antiparticles) in ${\bf J}_5$ are aligned to the direction
of ${\bf B}$ which can contribute to the hyperon polarization.  One
can test this by studying the dependence of the polarization on the
event charge asymmetry, $A_{\rm ch}=\langle N_{+}-N_{-}\rangle/\langle
N_{+}+N_{-}\rangle$ where $N_{+(-)}$ denotes the number of positively
(negatively) charged particles, assuming the relation $\mu_{\rm
  v}/T\propto A_{\rm ch}$.

\begin{figure}[htb]\vspace{0.1in}
\begin{center}
\includegraphics[width=\linewidth]{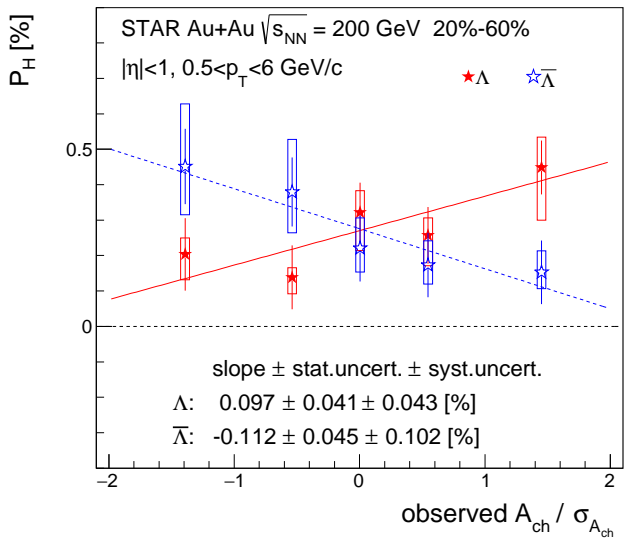}
\caption{Polarization of \lam and \alam as a function of observed
  charge asymmetry $A_{\rm ch}$ normalized with its RMS
  $\sigma_{A_{\rm ch}}$ for the 20\%--60\% centrality bin in Au+Au
  collisions at \sqsn = 200 GeV.  Open boxes and vertical lines show
  systematic and statistical uncertainties. Solid and dashed lines
  show linear fitting functions.\label{fig:PH_ChAsym}}
\end{center}
\end{figure}

Figure~\ref{fig:PH_ChAsym} presents the polarization as a function of
the event charge asymmetry $A_{\rm ch}$, where $A_{\rm ch}$ was
normalized by its RMS, $\sigma_{A_{\rm ch}}$, to avoid a possible
centrality bias, since the width of the $A_{\rm ch}$ distribution becomes
wider in peripheral collisions.  The results have large uncertainties,
but the dependence on $A_{\rm ch}/\sigma_{A_{\rm ch}}$ seems to be
different for \lam and \alam. The data were fitted with a
linear function and the extracted slope values are shown in
Fig.~\ref{fig:PH_ChAsym}.  The observed difference in slopes is a
 1--2$\sigma$ effect. If confirmed by higher statistics measurements, this observation 
 might open an important direction in studying chiral dynamics in heavy-ion collisions.

\section{Summary\label{sec:sum}}
We present the results of global polarization measurements for \lam
and \alam hyperons in Au+Au collisions at \sqsn = 200 GeV.  With a 150-fold 
improvement in statistics compared to the previous
measurements, we were able to measure the polarization with better
than a tenth of a percent accuracy. Depending on centrality, a non-zero
signal in the range of 0.1\%--0.5\% was observed. 
We find no significant difference between \lam and
\alam polarization at \sqsn = 200 GeV within the uncertainties.
The present global polarization measurement at 200 GeV with its relatively small uncertainty 
adds significance to the earlier observed trend at lower RHIC energies~\cite{polBES} of the global polarization decrease with the collision energy.
Within the uncertainties, our results agree with predictions from a
hydrodynamic (UrQMD+vHLLE) and the AMPT (A Multi-Phase
Transport) models. 

The polarization was also studied as functions of the collision
centrality, the hyperon's transverse momentum, and the pseudorapidity.  The
polarization was found to be larger in more peripheral
collisions, as expected from theoretical calculations, but no
significant dependence on pseudorapidity or transverse momentum was observed.
Furthermore, an indication of a polarization dependence on the
event-by-event charge asymmetry was observed. This might be an indication of a
possible contribution to the global polarization from the axial
current induced by the initial magnetic field, although the
statistical uncertainties need to be improved to reach a definitive
conclusion.
%
%
%
%
%
\begin{acknowledgments}
We thank the RHIC Operations Group and RCF at BNL, the NERSC Center at
LBNL, and the Open Science Grid consortium for providing resources and
support. This work was supported in part by the Office of Nuclear
Physics within the U.S. DOE Office of Science, the U.S. National
Science Foundation, the Ministry of Education and Science of the
Russian Federation, National Natural Science Foundation of China,
Chinese Academy of Science, the Ministry of Science and Technology of
China and the Chinese Ministry of Education, the National Research
Foundation of Korea, Czech Science Foundation and Ministry of
Education, Youth and Sports of the Czech Republic, Department of
Atomic Energy and Department of Science and Technology of the
Government of India; the National Science Centre of Poland, the
Ministry of Science, Education and Sports of the Republic of Croatia,
RosAtom of Russia and German Bundesministerium fur Bildung,
Wissenschaft, Forschung and Technologie (BMBF) and the Helmholtz
Association.
\end{acknowledgments}
%
\bibliography{ref_gpol}   
\end{document}